\documentclass[a4paper,twocolumn]{article}
\usepackage{amsfonts}

\usepackage{graphicx}
\usepackage{amsmath}

\newtheorem{proposition}{Proposition}
\begin{document}

\title{Entanglement of stabilizer codewords}
\author{Xiao-yu Chen, Li-zhen Jiang \\
{\small {College of Information and Electronic Engineering, Zhejiang
Gongshang University, Hangzhou, 310018, China}}}
\date{}
\maketitle

\begin{abstract}
The geometric measure, the logarithmic robustness and the relative entropy
of entanglement are proved to be equal for a stabilizer quantum codeword.
The entanglement upper and lower bounds are determined with the generators
of code. The entanglement of dual-containing CSS codes, Gottesman codes and
the related codes are given. An iterative algorithm is developed to
determine the exact value of the entanglement when the two bounds are not
equal.

\end{abstract}

\section{Introduction}

A quantum code encodes logical qubits in physical qubits. A quantum codeword
is usually a multipartite state if the physical qubits are owned by
different parties. The quantification of multipartite entanglement is
basically open even for a pure multipartite state until now. However, a
variety of different entanglement measures have been proposed for
multipartite setting. Among them are the \textit{(Global) Robustness of
Entanglement }\cite{Vidal}\textit{\ , }the \textit{Relative Entropy of
Entanglement}\cite{Vedral1} \cite{Vedral2}\textit{\ , }and \textit{the
Geometric Measure }\cite{Wei1}\textit{. } The robustness measures the
minimal noise (arbitrary state) that we need to add to make the state
separable. The geometric measure is the distance of the state to its closest
product state in terms of fidelity. The relative entropy of entanglement is
a valid entanglement measure for a multipartite state, it is the relative
entropy of the state to its closest separable state. The quantification of
multipartite entanglement is usually very difficult as most measures are
defined as the solutions to difficult variational problems. Even for pure
multipartite states, the entanglement can only be obtained for some special
scenarios. Fortunately, due to the inequality on the logarithmic robustness,
relative entropy of entanglement and geometric measure of entanglement\cite
{Wei2} \cite{Hayashi1} \cite{Wei3}, these entanglement measures are all
equal for stabilizer states \cite{Hayashi2}. A stabilizer state is a
multiqubit pure state which is the unique simultaneous eigenvector of a
complete set of commuting observables in the Pauli group, the latter
consisting of all tensor products of Pauli matrices and the identity with an
additional phase factor.

A stabilizer state is the special case of a quantum code that encodes zero
logical qubit. We may ask if the three entanglement measures are equal for a
generic quantum codeword. The answer is true as shown in Section II. The
three equal entanglement measures for a codeword then are simply called the
entanglement of the codeword. The rest of the paper is organized as follows.
In Section III, we derive the upper bound of the entanglement with Pauli
measurements. In Section IV, the lower bound of the entanglement is obtained
with the bipartition of the physical qubit system. Section V deals with the
entanglement of the codewords of CSS codes. The entanglement of the family
of Gottesman codes is the topic of Section VI. Section VII provides an
iterative method for the possible exact value of the entanglement when the
two bounds are not equal. Conclusions are drawn in Section VIII.

\section{The entanglement measure for quantum codewords}

The global robustness of entanglement $R(\rho )$ \cite{Vidal} is defined as
\begin{equation}
R(\rho )=\min t
\end{equation}
such that there exists a state $\Delta $, satisfying
\begin{equation}
\sigma =(\rho +t\Delta )/(1+t)\in Sep  \label{wave0}
\end{equation}
where $Sep$ is the set of separable states. The logarithmic robustness is
\begin{equation}
LR(\rho )=\log _2(1+R(\rho )).
\end{equation}
The relative entropy of entanglement is defined as the ''distance'' to the
closest separable state in terms of relative entropy \cite{Vedral2},
\begin{equation}
E_r(\rho )=\min_{\omega \in Sep}S\left( \rho \right\| \left. \omega \right) ,
\end{equation}
where $S\left( \rho \right\| \left. \omega \right) =-S(\rho )-$ $tr\{\rho
log_2\omega \}$ is the relative entropy, $S(\rho )$ is the von Neumann
entropy.The geometric measure of entanglement for pure state $\left| \psi
\right\rangle $, is defined as
\begin{equation}
E_g(\left| \psi \right\rangle )=\min_{\left| \phi \right\rangle \in \mathit{%
Pro}}-\log _2\left| \left\langle \phi \right| \left. \psi \right\rangle
\right| ^2,  \label{wee0}
\end{equation}
where $Pro$ is the set of product states. An extension of the definition for
mixed state $\rho $ is also available, $E_g(\rho )=\min_{\omega \in
Sep}-\log _2tr(\rho \omega )$, however, $E_g$ is an entanglement monotone
only for pure states $\rho =\left| \psi \right\rangle \left\langle \psi
\right| .$ It has been shown that the maximal number $N$ of pure states in
the set $\{\left| \psi _i\right\rangle |i=1,...,N\}$, that can be
discriminated perfectly by LOCC is bounded by the amount of entanglement
they contain\cite{Hayashi1}:
\begin{equation}
\log _2N\leq n-\overline{LR(\left| \psi _i\right\rangle )}\leq n-\overline{%
E_r(\left| \psi _i\right\rangle )}\leq n-\overline{E_g(\left| \psi
_i\right\rangle )},  \label{wee}
\end{equation}
where $n=\log _2D_H,$ $D_H$ is the total dimension of the Hilbert space, and
$\overline{x}=\frac 1N\sum_{i=1}^Nx_i$ denotes the ''average''.

An $n$-qubit stabilizer state $\left| S\right\rangle $ is defined as a
simultaneous eigenvector with eigenvalue $1$ of $n$ commuting and
independent Pauli group elements $M_i$. The $n$ eigenvalue equations $%
M_i\left| S\right\rangle =\left| S\right\rangle $ define the state $\left|
S\right\rangle $ completely (up to an arbitrary phase). The group generated
by the product of the $n$ operators $M_i$ is called the stabilizer $S,$ and $%
M_i$ are generators of $S.$ A subgroup of $S$ with $n-k$ generators is also
called a stabilizer\cite{Gottesman97}, denoted as $M.$ However, $M\subset S$
stabilizes a $2^k$ dimensional space. In principle, such a space is the
coding space $\{\left| \psi \right\rangle ,s.t.T\left| \psi \right\rangle
=\left| \psi \right\rangle ,\forall T\in M\}$, corresponds to a stabilizer
code encoding $k$ into $n$ qubits. In addition to the $n-k$ stabilizer
generators, a stabilizer code also has logical operations $\overline{X}%
_1,\ldots ,\overline{X}_k$ and $\overline{Z}_1,\ldots ,\overline{Z}_k.$ We
can take the basis codewords for this code to be
\begin{eqnarray*}
\left| \overline{0}\right\rangle &=&\prod_{T\in M}^{}T\left| 0\right\rangle
^{\otimes n}, \\
\left| \overline{\mathbf{c}}\right\rangle &=&\overline{X}_1^{c_1}\cdots
\overline{X}_k^{c_k}\left| \overline{0}\right\rangle ,
\end{eqnarray*}
where $\mathbf{c=}(c_1,\ldots ,c_k)$ is a binary vector. $\overline{Z}%
_i\left| \overline{0}\right\rangle =\left| \overline{0}\right\rangle $ for $%
i=1,\ldots ,k.$

For a stabilizer state $\left| S\right\rangle $, it has been shown that\cite
{Hayashi2}
\begin{equation}
LR(\left| S\right\rangle )=E_r(\left| S\right\rangle )=E_g(\left|
S\right\rangle ).
\end{equation}
This is also true for a stabilizer quantum codeword $\left| \overline{%
\mathbf{c}}\right\rangle $.

\begin{proposition}
The entanglement measures of the logarithmic robustness, the relative
entropy of entanglement and the geometric measure of entanglement are all
equal for a stabilizer quantum codeword $\left| \overline{\mathbf{c}}%
\right\rangle .$
\end{proposition}

Proof: It is enough to prove the statement for $\left| \overline{0}%
\right\rangle ,$ since $\left| \overline{\mathbf{c}}\right\rangle $ is
locally equivalent to $\left| \overline{0}\right\rangle .$ Notice that $%
\left| \overline{0}\right\rangle $ is stabilized by $\{M_1,\ldots ,M_{n-k},%
\overline{Z}_1,\ldots ,\overline{Z}_k\}.$ The $n-k$ generators and $k$
logical $\overline{Z}$ operators commute with each other and are
independent. Thus $\left| \overline{0}\right\rangle $ is a stabilizer state,
the three entanglement measures are equal according to Ref. \cite{Hayashi2}.

For a stabilizer quantum codeword, we will simply call these three
quantities the entanglement, and denote as $E(\left| \overline{\mathbf{c}}%
\right\rangle ).$

\section{Entanglement upper bound}

A generator which is composed of $Z$ operator or identity in each qubit and
does not contain $X$ or $Y$ operators in any qubit is called a Z-type
generator.

\begin{proposition}
\label{proposition2} The entanglement of a codeword is upper bounded by the
minimal number of stabilizer generators which are not Z-type generators.
\end{proposition}

Proof: The codeword $\left| \overline{0}\right\rangle
=N\prod_{i=1}^{n-k}(I+M_i)\left| 0\right\rangle ^{\otimes n},$where $N$ is
the normalization factor. For a Z-type generator $M_l$, we have $%
(I+M_l)\left| 0\right\rangle ^{\otimes n}=2\left| 0\right\rangle ^{\otimes
n}.$ In the operator product $\prod_i(I+M_i)$, we may move the factor $%
(I+M_l)$ to the rightmost. The number of terms $R$ in the linear
decompositions of $\left| \overline{0}\right\rangle $ into product states is
upper bounded by $2^{r^{\prime }},$ with $r^{\prime }$ the number of non
Z-type generators. Let $r=\min r^{\prime }$ be the minimal number of non
Z-type generators. Notice that we can artificially increase the number of
non Z-type generators by replacing a Z-type generator with the product of
the Z-type generator and a non Z-type generator. Hence we count the minimal
number of non Z-type generators. It follows that the Schmidt measure \cite
{Hein}
\[
E_s=\min \log _2R
\]
is upper bounded by $r.$ The geometric measure is upper bounded by Schmidt
measure \cite{Markham10}, hence the theorem follows.

For a stabilizer group $M$, up to over all phase $\pm 1,\pm i,$ each
generator $M_i=X^{a_i}Z^{b_i},$ with $X^{a_i}=%
\bigotimes_jX_j^{a_{ij}},Z^{b_i}=\bigotimes_jZ_j^{b_{ij}}$, where $a_i$ and $%
b_i$ are the binary vectors $(a_{i1},a_{i2},\ldots ,a_{in})$ and $%
(b_{i1},b_{i2},\ldots ,b_{in}),$ respectively. An alternative representation
of the generator $M_i$ is $\left( a_i\right| \left. b_i\right) .$ We may use
generator matrix ( follow Ref. \cite{Hein06}, it is called stabilizer matrix
in Ref. \cite{Gottesman97} ) $\left( A\right| \left. B\right) $ to represent
the stabilizer group $M,$ where $A$ and $B$ are $(n-k)\times n$ matrices
with elements $A_{ij}=a_{ij},$ $B_{ij}=b_{ij}.$It is always possible to
arrange $(A|B)$ in the form of (see e.g. \cite{Gottesman97} Ch.4)
\begin{equation}
\left(
\begin{array}{ll}
I & D \\
0 & 0
\end{array}
\right. \left|
\begin{array}{ll}
F & G \\
J & K
\end{array}
\right)  \label{wave3a}
\end{equation}
by the permutations of the qubits and replacements of the generators with
other elements in the stabilizer group. Here $I$ is an $r\times r$ identity
matrix, with $r$ the $\Bbb{F}_2$ rank of $A,$ where $\Bbb{F}_2$ denotes the
integer field $\{0,1\}$ with addition and multiplication modulo 2. With the
standard form (\ref{wave3a}) of the generator matrix, we may improve the
upper bound of the entanglement. We then investigate the effect of Pauli
measurements on codewords. We work on the generators of $M$ in the first $r$
lines of (\ref{wave3a}) and neglect the other $n-k-r$ Z-type generators. We
have $M_1=X\otimes M_1^{\prime }$ or $M_1=Y\otimes M_1^{\prime },$ $%
M_{i+1}=Z\otimes M_{i+1}^{\prime }$or $M_{i+1}=I\otimes M_{i+1}^{\prime }$ ($%
i=1,\ldots ,r-1$) for the standard form of the generators. Denote $\left|
\overline{0}_{n-1}\right\rangle =N^{\prime
}\prod_{i=1}^{r-1}(I+M_{i+1}^{\prime })\left| 0\right\rangle ^{\otimes
(n-1)},$where $N^{\prime }$ is the normalization factor, then Pauli $Z$
measurement on the first qubit will project the codeword $\left| \overline{0}%
\right\rangle $ to
\begin{eqnarray}
P_{z+}^{(1)}\left| \overline{0}\right\rangle &=&\left| 0\right\rangle
\otimes \left| \overline{0}_{n-1}\right\rangle ,  \label{wave4} \\
P_{z-}^{(1)}\left| \overline{0}\right\rangle &=&\left| 1\right\rangle
\otimes M_1^{\prime }\left| \overline{0}_{n-1}\right\rangle .  \label{wave5}
\end{eqnarray}
The projection operators on the $j^{th}$ qubit are $P_{z\pm }^{(j)}=\frac
12(I\pm Z_j).$ The two measurement results $\pm 1$ are equally probable.
Similarly, the $X$ or $Y$ measurements on the first qubit also project the
codeword to two equally probable states corresponding to the two measurement
results $\pm 1,$ except for possible special case of $X$ or $Y$ measurements
with only one result$.$ When $M_1=X\otimes M_1^{\prime },$ the Pauli $X,Y$
measurements on the first qubit will project the codeword $\left| \overline{0%
}\right\rangle $ to $P_{x\pm }^{(1)}\left| \overline{0}\right\rangle $ $%
=\frac 12(\left| 0\right\rangle \pm \left| 1\right\rangle )$ $\otimes (I\pm
M_1^{\prime })\left| \overline{0}_{n-1}\right\rangle ,$ $P_{y\pm
}^{(1)}\left| \overline{0}\right\rangle $ $=\frac 12(\left| 0\right\rangle
\pm i\left| 1\right\rangle )$ $\otimes (I\mp iM_1^{\prime })\left| \overline{%
0}_{n-1}\right\rangle .$ When $M_1=Y\otimes M_1^{\prime },$ the Pauli $X,Y$
measurements on the first qubit will project the codeword $\left| \overline{0%
}\right\rangle $ to $P_{x\pm }^{(1)}\left| \overline{0}\right\rangle $ $%
=\frac 12(\left| 0\right\rangle \pm \left| 1\right\rangle )$ $\otimes (I\pm
iM_1^{\prime })\left| \overline{0}_{n-1}\right\rangle ,$ $P_{y\pm
}^{(1)}\left| \overline{0}\right\rangle $ $=\frac 12(\left| 0\right\rangle
\pm i\left| 1\right\rangle )$ $\otimes (I\pm M_1^{\prime })\left| \overline{0%
}_{n-1}\right\rangle .$ There is the case that $\left| \overline{0}%
_{n-1}\right\rangle $ is the eigenvector of $M_1^{\prime }$, so occasionally
one of the two outcomes is annihilated.

As a property of the Schmidt measure \cite{Hein}, for any sequence of local
projective measurements that finally completely disentangles the state
vector $\left| \psi \right\rangle $ in each of the measurement results, we
obtain the upper bound

\begin{equation}
E_S(\left| \psi \right\rangle \leq \log _2(N_{mea}),  \label{wave5a}
\end{equation}
where $N_{mea}$ is the number of measurement results with non-zero
probability.

The minimal number of local Pauli measurements to disentangle a stabilizer
quantum codeword can be called its ''Pauli persistency''.

\begin{proposition}
\label{proposition3} The entanglement is upper bounded by '' Pauli
persistency'' for a stabilizer quantum codeword $\left| \overline{\mathbf{c}}%
\right\rangle .$
\end{proposition}

Proof: With the formula (\ref{wave5a}), the fact that different measurement
results of codeword are obtained with probability $1/2$, and geometric
measure is upper bounded by Schmidt measure, the statement follows.

\textit{Example.---}Consider\textit{\ }$[[8,1,3]]$ code \cite{Grassl} with
stabilizer in the standard form
\[
\begin{array}{llllllll}
X & Z & Z & Z & Z & Z & Z & Y \\
I & X & I & Z & I & Z & I & Z \\
I & I & X & Z & I & I & Z & Z \\
Z & Z & Z & X & I & I & I & I \\
I & Z & Z & I & Y & Z & Z & X \\
Z & Z & I & I & Z & X & I & I \\
Z & Z & I & I & I & I & X & Y
\end{array}
,
\]
the Pauli $Z$ measures are applied at the $1,5,7$ qubits. For each qubit
measured, the corresponding row and column are deleted. What left for the
remain qubits is the stabilizer
\begin{equation}
\begin{array}{lllll}
X & I & Z & Z & Z \\
I & X & Z & I & Z \\
Z & Z & X & I & I \\
Z & I & I & X & I
\end{array}
.  \label{wave5b}
\end{equation}
The $\left| \overline{0}_5\right\rangle $ generated by stabilizer (\ref
{wave5b}) is a product of graph state $\left| G_4\right\rangle $ with $%
\left| 0\right\rangle ,\left| \overline{0}_5\right\rangle =\left|
G_4\right\rangle $ $\otimes $ $\left| 0\right\rangle .$ The graph state $%
\left| G_4\right\rangle $ is generated by a new stabilizer obtained with
deleting the last column of (\ref{wave5b}). ''Pauli persistency'' of graph
state $\left| G_4\right\rangle $ is $2,$ so ''Pauli persistency'' for
codeword of $[[8,1,3]]$ code in Grassl code-table \cite{Grassl} is $5.$

We will use proposition \ref{proposition2} to obtain the entanglement upper
bound of codeword in the following except Table 1, where proposition \ref
{proposition3} is used for tighter upper bounds.

\section{Entanglement lower bound \label{sec4}}

The index of physical qubits is denoted as $\mathcal{I}=\{1,2,\ldots ,n\},$
for a bipartition, we may assign $m$ qubits to $\mathcal{A}$, and the remain
$n-m$ qubits to $\mathcal{B}$. The index sets of $\mathcal{A}$ and $\mathcal{%
B}$ are $\mathcal{I}_{\mathcal{A}}$ and $\mathcal{I}_{\mathcal{B}}=\mathcal{%
I-}$ $\mathcal{I}_{\mathcal{A}},$ respectively. The reduced state of the
codeword $\left| \overline{0}\right\rangle $ then should be $\rho _{\mathcal{%
B}}=Tr_{\mathcal{A}}\left| \overline{0}\right\rangle \left\langle \overline{0%
}\right| .$ The bipartite entanglement of the bipartition$\{\mathcal{I}_{%
\mathcal{A}},\mathcal{I}_{\mathcal{B}}\}$ then will be $-Tr\rho _{\mathcal{B}%
}\log _2\rho _{\mathcal{B}},$ the entropy of $\rho _{\mathcal{B}}.$

\begin{proposition}
The entanglement of a codeword is lower bounded by any bipartite
entanglement of the codeword,
\begin{equation}
E\geq -Tr\rho _{\mathcal{B}}\log _2\rho _{\mathcal{B}}.  \label{wave5c}
\end{equation}
\end{proposition}

Proof: If we define $E_{rbi}$ as the relative entropy of entanglement with
respect to some bipartition, we have that $E_r\geq E_{rbi}$ since the set of
fully separable states is a subset of the bipartite separable states. Notice
that $E_r=E$ and $E_{rbi}$ is equal to the bipartite entanglement $%
E_{bi}=-Tr\rho _{\mathcal{B}}\log _2\rho _{\mathcal{B}}$ for pure state, the
statement then follows.

We will obtain the entropy of $\rho _{\mathcal{B}}$ by
diagonalizing $\rho _{\mathcal{B}}$ and at last the entropy can be
expressed with the code stabilizer. The entanglement of the
codeword is lower bounded by the maximal bipartite entanglement
among all bipartitions.

Since a Z-type generator does not contribute new items to codeword $\left|
\overline{0}\right\rangle ,$ we simply ignore Z-type generators. Thus we
take $\left( A\right. \left| B\right) =\left( I\right. D\left| E\text{ }%
F\right) $ in the following. $A,B$ are $r\times n$ binary matrices with $r$
the number of non Z-type generators and $r\leq n-k.$

The codeword
\begin{eqnarray*}
\left| \overline{0}\right\rangle &=&N\sum_\mu (-1)^{\alpha (\mu )}X^{\mu
A}\left| 0\right\rangle ^{\otimes n} \\
&=&N\sum_\mu (-1)^{\alpha (\mu )}\bigotimes_{j\in \mathcal{I}_{\mathcal{A}%
}}X_j^{(\mu A)_j}\bigotimes_{l\in \mathcal{I}_{\mathcal{B}}}X_l^{(\mu
A)_l}\left| 0\right\rangle ^{\otimes n},
\end{eqnarray*}
where the summation on $r$ dimensional binary vector $\mu $ is from $%
(0,0,\ldots ,0)$ to $(1,1,\ldots 1),$ $(\mu A)_j$ is the $j^{th}$ component
of the binary vector $\mu A,$ $N$ is the normalization factor, and $\alpha
(\mu )=\sum_{i<l}(\mu _ia_i)(\mu _lb_l^T).$ We my rewrite $\alpha (\mu )$ as
\begin{equation}
\alpha (\mu )=\frac 12[\mu \Gamma \mu ^T-Tr(\Lambda \Gamma \Lambda )]=\frac
12\mu \Gamma _1\mu ^T,
\end{equation}
with $\Lambda =diag\{\mu _1,\ldots ,\mu _r\},$ and $\Gamma _1$ is the matrix
$\Gamma $ with diagonal elements nullified, where
\[
\Gamma =AB^T=F^T+DG^T.
\]
The convention for binary addition is mod 2. $\Gamma $ is symmetric for any
two generators should commute with each other, namely, $AB^T+B^TA=0.$ The
reduced state $\rho _{\mathcal{B}}=\sum_{\mu ,\mu ^{\prime }}\prod_{j\in
\mathcal{I}_{\mathcal{A}}}\delta _{(\mu A)_j,(\mu ^{\prime
}A)_j}(-1)^{\alpha (\mu )+\alpha (\mu ^{\prime })}$ $\bigotimes_{l,l^{\prime
}\in \mathcal{I}_{\mathcal{B}}}X_l^{(\mu A)_l}\left| 0\right\rangle
^{\otimes (n-m)}$ $\left\langle 0\right| ^{\otimes (n-m)}$ $X_{l^{\prime
}}^{(\mu ^{\prime }A)_{l^{\prime }}}$ disregarding normalization, which is
\[
\rho _{\mathcal{B}}=\sum_{\mu ,\mu ^{\prime }}\prod_{j\in \mathcal{I}_{%
\mathcal{A}}}\delta _{(\mu A)_j,(\mu ^{\prime }A)_j}(-1)^{\alpha (\mu
)+\alpha (\mu ^{\prime })}\left| (\mu A)_{\mathcal{B}}\right\rangle
\left\langle (\mu A)_{\mathcal{B}}\right| ,
\]
with $\left| (\mu A)_{\mathcal{B}}\right\rangle =\left| (\mu A)_{\mathcal{I}%
_{m+1}},\ldots (\mu A)_{\mathcal{I}_n}\right\rangle .$

We then consider to diagonalize $\rho _{\mathcal{B}}$ in order to obtain its
entropy. Without loss of generality, let $\mathcal{I}_1=1,$ $\mathcal{I}%
_2=2, $ $\mathcal{I}_m=m\leq r,$ and denote $\mu =(\nu ,\tau ),$with $\nu
=(\mu _1,\ldots ,\mu _m),\tau =(\mu _{m+1},\ldots ,\mu _r).$ then$\left|
(\mu A)_{\mathcal{B}}\right\rangle $ $=$ $\left| \mu _{m+1},\ldots ,\mu
_r,(\mu D)_1,\ldots ,(\mu D)_{n-r}\right\rangle =\left| \tau ,\mu
D\right\rangle .$ Denote $\left| \Psi (\nu )\right\rangle =$ $\sum_\tau
(-1)^{\alpha (\mu )}\left| (\mu A)_{\mathcal{B}}\right\rangle ,$ then
\begin{eqnarray*}
\rho _{\mathcal{B}} &=&\sum_{\nu ,\nu ^{\prime },\tau ,\tau ^{\prime
}}\delta _{\nu ,\nu ^{\prime }}(-1)^{\alpha (\mu )+\alpha (\mu ^{\prime
})}\left| (\mu A)_{\mathcal{B}}\right\rangle \left\langle (\mu A)_{\mathcal{B%
}}\right| \\
&=&\sum_{\nu ,\nu ^{\prime }}\delta _{\nu ,\nu ^{\prime }}\left| \Psi (\nu
)\right\rangle \left\langle \Psi (\nu ^{\prime })\right| =\sum_\nu \left|
\Psi (\nu )\right\rangle \left\langle \Psi (\nu )\right| .
\end{eqnarray*}
For the orthogonality of $\left| \Psi (\nu )\right\rangle $ , we turn to
\begin{eqnarray}
\left\langle \Psi (\nu ^{\prime })\right. \left| \Psi (\nu )\right\rangle
=\sum_{\tau ,\tau ^{\prime }}(-1)^{\alpha (\mu )+\alpha (\mu ^{\prime
})}\delta _{\tau \tau ^{\prime }}\delta _{\mu D,\mu ^{\prime }D}  \nonumber
\\
=\sum_\tau (-1)^{\alpha (\nu ,\tau )+\alpha (\nu ^{\prime },\tau )}\delta
_{(\nu ,\tau )D,(\nu ^{\prime },\tau )D}.  \label{wave6}
\end{eqnarray}
The bipartite entanglement of the codeword is at least $m$ when all $\left|
\Psi (\nu )\right\rangle $ are orthogonal with each other. This is obviously
from the factor that
\[
\rho _{\mathcal{B}}=\frac 1{2^m}\sum_{\nu =(0,0,\ldots ,0)}^{(1,1,\ldots
,1)}\left| \Psi (\nu )\right\rangle \left\langle \Psi (\nu )\right| ,
\]
where $\left| \Psi (\nu )\right\rangle $ are orthonormal and the
normalization factor is retrieved. The bipartite entanglement may be less
than $m$ only when the non-orthogonality of the ensemble $\left| \Psi (\nu
)\right\rangle $ is found. The conditions for nonzero $\left\langle \Psi
(\nu ^{\prime })\right. \left| \Psi (\nu )\right\rangle $ are
\begin{eqnarray}
(\nu +\nu ^{\prime },\mathbf{0})D &=&0,  \label{wave10} \\
(\nu +\nu ^{\prime })\Gamma _3 &=&0,  \label{wave11}
\end{eqnarray}
where $\mathbf{0}$ stands for the $r-m$ dimensional zero vector $(0,0,\ldots
,0)$. $\Gamma _3$ is produced by deleting the first $m$ columns and the last
$r-m$ rows of the $r\times r$ matrix $\Gamma _1$, so $\Gamma _3$ is a $%
m\times (r-m)$ submatrix of $\Gamma _1$. More explicitly, we may write $%
\Gamma _1$ as
\[
\Gamma _1=\left[
\begin{array}{ll}
\Gamma _2 & \Gamma _3 \\
\Gamma _3^T & \Gamma _4
\end{array}
\right] .
\]
Then $\alpha (\nu ,\tau )+\alpha (\nu ^{\prime },\tau )=\frac 12(\nu \Gamma
_2\nu ^T+\nu ^{\prime }\Gamma _2\nu ^{\prime T})+(\nu +\nu ^{\prime })\Gamma
_3\tau ^T+\tau \Gamma _4\tau ^T.$ The $\Gamma _4$ term always contributes a $%
+1$ factor in the summation of Eq.(\ref{wave6}) for $\Gamma _4$ is symmetric
and with nullified diagonal elements. The $\Gamma _2$ term contributes a
constant factor in the summation of Eq.(\ref{wave6}). Then Eq.(\ref{wave11})
follows. Let $D^{\prime }$ be the matrix produced by deleting the last $r-m$
rows and preserving the first $m$ rows of $D$, the rank of the $m\times
(n-m) $ matrix $Q(\mathcal{A},\mathcal{B})=(\Gamma _3,D^{\prime })$ gives
the number of independent vectors $\left| \Psi (\nu )\right\rangle $ for a
specific bipartition of first $m$ qubits for $\mathcal{A}$ with respect to
last $n-m$ qubits for $\mathcal{B}$. Maximizing with respect to all
bipartitions except the $m>r$ cases, we hence obtain the maximal of
bipartite entanglement as the lower bound of the entanglement
\begin{equation}
E_l=\max_{partitions}rank_{\Bbb{F}_2}Q(\mathcal{A},\mathcal{B}).
\end{equation}
Since $Q(\mathcal{A},\mathcal{B})$ is a $m\times (n-m)$ matrix, its rank
must not exceed $\min \{m,n-m\}\leq \left\lfloor \frac n2\right\rfloor ,$ so
we have $E_l\leq \left\lfloor \frac n2\right\rfloor .$

One of the special case that should be notified is when $m=r.$ It follows
that $\Gamma _3$ is an $r\times 0$ matrix and does not exist at all. So Eq.(%
\ref{wave11}) disappears, and we only need to consider Eq.(\ref{wave10}). If
$rank_{\Bbb{F}_2}D=r$, then Eq.(\ref{wave10}) fulfills only when $%
\nu=\nu^{\prime}$, hence $E_l=r.$

\section{Entanglement of CSS codes}

\subsection{Dual-containing CSS codes}

An important class of quantum codes, constructed from classical codes,
invented by Calderbank, Shor \cite{Calderbank} \& Steane \cite{Steane}, has
the generator matrix of the form (e.g. \cite{MacKay})
\begin{equation}
(A|B)=\left(
\begin{array}{l}
U \\
0
\end{array}
\right. \left|
\begin{array}{l}
0 \\
V
\end{array}
\right) ,  \label{wav}
\end{equation}
where $U$ and $V$ are $l\times n$ matrices. Requiring $UV^T=0$ ensures that
the generators commute with each other. As there are $2l$ stabilizer
conditions applying to $n$ qubit states, $k=n-2l$ qubits are encoded in $n$
qubits. We may write the classical parity check matrix $U$ in a systematical
way
\begin{equation}
U=\left[
\begin{array}{ll}
I & D
\end{array}
\right] .  \label{wav1}
\end{equation}
Since $AB^T=0\mathbf{,}$ we get $\alpha (\mu )=0$ for all binary vectors $%
\mu .$ Consider the case of $m=l,$ we have $\nu =\mu $. Then $\left\langle
\Psi (\nu ^{\prime })\right. \left| \Psi (\nu )\right\rangle =\delta _{\nu
D,\nu ^{\prime }D}$. From $\nu D=\nu ^{\prime }D,$ we have $(\nu +\nu
^{\prime })D=0.$ Thus the condition for the orthogonality of $\left| \Psi
(\nu )\right\rangle $ is
\begin{equation}
\nu D=0\text{ }\Rightarrow \text{ }\nu =0,  \label{wave7}
\end{equation}
for all binary vector $\nu .$ The lower bound of the entanglement of the
codeword is $l$ when the condition (\ref{wave7}) is fulfilled. For
dual-containing code, we have $V=U,$ thus $UU^T=0,$ so that $DD^T=I,$ the
condition (\ref{wave7}) is fulfilled. The lower bound of entanglement is $%
E_l=l=\frac{n-k}2$

The upper bound of the entanglement $E_u$ of the codeword $\left| \overline{0%
}\right\rangle $ is the number of $X$ generators now, which is $l.$ Thus the
entanglement of dual-containing CSS codeword is
\begin{equation}
E=\frac{n-k}2.  \label{wave7a}
\end{equation}
For a CSS code that is not dual-containing, the upper bound $E_u$ is still $%
l $, the number of $X$ generators. The lower bound is the binary rank of $D.$

\subsection{The graph state of a CSS code}

A stabilizer code with stabilizer generators $M_1,\ldots ,M_{n-k}$ and
logical operations $\overline{X}_1,\ldots ,\overline{X}_k$ and $\overline{Z}%
_1,\ldots ,\overline{Z}_k$, is equivalent to the codeword stabilizer (CWS)
code \cite{Cross} defined by codeword stabilizer $\{M_1,\ldots ,M_{n-k},%
\overline{Z}_1,\ldots ,\overline{Z}_k\}$ and word operators which are
products of $\overline{X}_i$. Any CWS code is locally Clifford-equivalent to
a standard form of CWS code with a graph-state stabilizer and word operators
consisting only of $Z$ operators. The standard codeword stabilizer is
generated by $X_iZ^{\mathbf{r}_i}$ . The set of $\mathbf{r}_is$ forms the
adjacency matrix of the graph \cite{Cross}. Hence, given a quantum
stabilizer error-correcting code, we can always find the corresponding graph
state. The entanglement of the codeword of the quantum stabilizer code and
the graph state should be the same, since they are locally
Clifford-equivalent. A CSS code has a generator matrix (\ref{wav}) and $U$
can further written in the form of (\ref{wav1}). We now construct the graph
state stabilizer. The generator matrix of $\{M_1,\ldots ,M_{n-k},\overline{Z}%
_1,\ldots ,\overline{Z}_k\}$ is
\begin{equation}
\left(
\begin{array}{l}
U \\
0 \\
0
\end{array}
\right. \left|
\begin{array}{l}
0 \\
V \\
W
\end{array}
\right) ,
\end{equation}
where $\left( 0\right. \left| W\right) $ is the generator matrix for logical
operations $\overline{Z}_1,\ldots ,\overline{Z}_k.$ With elementary row
transformation we have transformed $U\ $ into the systematical form $\left[
\begin{array}{ll}
I & D
\end{array}
\right] .$ We want show that it is always possible to transform $\left[
\begin{array}{l}
V \\
W
\end{array}
\right] $ into the form of $\left[
\begin{array}{ll}
D^{\prime } & I
\end{array}
\right] ,$ where $I$ is an $(n-l)\times (n-l)$ identity matrix. We first
transform $\left[
\begin{array}{l}
V \\
W
\end{array}
\right] $ into $\left[
\begin{array}{ll}
R & P
\end{array}
\right] ,$ where $P$ is an upper triangle square matrix, namely $P_{ij}=0$
for $i>j$. There is the case that $P_{jj}=0,$ we then interchange the $j-th$
qubit with some later qubit such that $P_{jj}=1.$ This is always possible
since the elements of the $j-th$ line of $P$ can not be all zeros, otherwise
the elements of $j-th$ line of $\left[
\begin{array}{ll}
R & P
\end{array}
\right] $ should be all zeros due to mutual commutation of the generators,
namely, $UV^T=0$ and $UW^T=0$. That is
\begin{equation}
IR^T+DP^T=0.  \label{wave7b}
\end{equation}
From which we can deduce that if the $j-th$ line of $P$ are all zeros, we
have $R_{ji}=0,$ for all $i\leq l.$ An all zero line in $\left[
\begin{array}{ll}
R & P
\end{array}
\right] $ means the generator is the identity, this is not the case. It is
easy to transform $\left[
\begin{array}{ll}
R & P
\end{array}
\right] $ to $\left[
\begin{array}{ll}
D^{\prime } & I
\end{array}
\right] $, then Eq. (\ref{wave7b}) will be
\[
D^{\prime }=D^T.
\]
Performing Hadamard transformation to the last $n-l$ qubits, the generator
matrix undergoes the transformation
\[
\left(
\begin{array}{ll}
I & D \\
0 & 0
\end{array}
\right. \left|
\begin{array}{ll}
0 & 0 \\
D^T & I
\end{array}
\right) \Rightarrow \left(
\begin{array}{ll}
I & 0 \\
0 & I
\end{array}
\right. \left|
\begin{array}{ll}
0 & D \\
D^T & 0
\end{array}
\right) .
\]
Thus the adjacency matrix of the locally Clifford-equivalent graph state of
CSS codeword is
\begin{equation}
\gamma =\left[
\begin{array}{ll}
0 & D \\
D^T & 0
\end{array}
\right] .  \label{wav2}
\end{equation}

A graph state with adjacency matrix of (\ref{wav2}) is two-colorable, we can
simply assign the first $l$ qubits with one color and the remain $n-l$
qubits with another. The entanglement upper bound should be \cite{Markham} $%
E_u=n-(n-l)=l.$ The lower bipartite bound is \cite{Hein} $E_l=\frac 12rank_{%
\mathbf{F}_2}\gamma =rank_{\mathbf{F}_2}D.$

For a dual-containing CSS code, the corresponding graph state is further
characterized by $DD^T=I$ in addition to two-colorable. The $D$ matrix has a
full rank and the upper and lower bounds of entanglement coincide. The
entanglement of the codeword is also $E=l$ with the theory of graph state
and formula (\ref{wav2}).

\subsection{Toric Codes}

Toric code is proposed to encode quantum information in topological
structure \cite{Kiteav}, it is a kind of quantum LDPC code \cite{MacKay}.
The toric code is based on a $k\times k$ square lattice on the torus. Each
edge of the lattice is attached with a qubit, so there are $n=2k^2$ qubits.
For each vertex $s$ and each face $p$, operators of the following form are
defined:

\[
A_s=\prod_{j\in star(s)}X_j,\text{ }B_p=\prod_{j\in boundary(p)}Z_j.
\]
These operators commute with each other. Due to $\prod_sA_s$ $=1$ and $%
\prod_pB_p$ $=1$, there are $m=2k^2-2$ independent operators constitute the
stabilizer of the toric code. The code encodes $n-m=2$ qubits. Toric code is
a kind of CSS code from its definition. We will show that it is not a dual
containing code, the upper and lower bound of entanglement may not coincide.
By proper numbering the edges, the generator matrix can be written in the
following form
\begin{eqnarray*}
U &=&\left[
\begin{array}{llllllllll}
I &  &  &  & I & \Omega ^{} &  &  &  &  \\
I & I &  &  &  &  & \Omega &  &  &  \\
& \ddots & \ddots &  &  &  &  & \ddots &  &  \\
&  & I & I &  &  &  &  & \Omega ^{} &  \\
&  &  & I^{\prime } & I^{\prime } &  &  &  &  & \Omega ^{\prime }
\end{array}
\right] , \\
V &=&\left[
\begin{array}{llllllllll}
\Omega ^T &  &  &  &  & I &  &  &  & I \\
& \Omega ^T &  &  &  & I & I &  &  &  \\
&  & \ddots &  &  &  & \ddots & \ddots &  &  \\
&  &  & \Omega ^T &  &  &  & I & I &  \\
&  &  &  & \Omega ^{T\prime } &  &  &  & I^{\prime } & I^{\prime }
\end{array}
\right] .
\end{eqnarray*}
where $I$ is the $k\times k$ identity matrix, $\Omega $ is a $k\times k$
matrix with $2k$ nonzero entries,
\[
\Omega =\left[
\begin{array}{llll}
1 &  &  & 1 \\
1 & 1 &  &  \\
& \ddots & \ddots &  \\
&  & 1 & 1
\end{array}
\right] ,
\]
where $I^{\prime }$ , $\Omega ^{\prime }$ and $\Omega ^{T\prime }$are the
matrices obtained by deleting the last line of $I$, $\Omega $ and $\Omega
^T, $ respectively. In order to show that the $D$ matrix has not a full rank
in general, we transform it into the following form by elementary row
transformation:
\[
\left[
\begin{array}{llllll}
I &  &  &  & \Omega &  \\
& \Omega &  &  &  &  \\
&  & \ddots &  &  &  \\
&  &  & \Omega &  &  \\
0 & \Delta & \cdots & \Delta & \Delta & \mathbf{1}
\end{array}
\right] ,
\]
where $\mathbf{1=}(1,1,\ldots ,1)^T$ is a $k-1$ dimensional column vector, $%
\Delta =[I_{k-1},\mathbf{1].}$ The $(k^2-1)\times (k^2+1)$ matrix $D$ is
apparently not a full rank matrix, for the rank of $\Omega $ is $k-1,$ we
have the lower bound of entanglement
\[
E_l=k^2-k+1.
\]
While the upper bound of entanglement is $E_u=k^2-1.$ Hence the upper bound
is no longer equal to the lower bound unless $k=2.$

The codeword of a toric code is a highly entangled state for large $k,$ the
entanglement scales as $E\sim \frac n2$. It seems that the area law \cite
{Eisert} does not work for toric code. Area law usually means that the
entanglement is proportional to the boundary area when the bulk of qubits is
cut to two parts. The number of bulk qubits now is $n=2k^2,$ so according to
area law, the (bipartite) entanglement should be proportional to the cutting
length $L,$ which is now proportional to $k$ in most cases. However, the
largest bipartite entanglement is proportional to $k^2,$ corresponding to a
sophisticated cutting curve with length $L\ \propto k^2$ , despite that a
random cutting of toric qubits into two parts usually yields a boundary
length $L\ \propto \sqrt{n}.$ Hence the area law still works, but the
boundary may be very long. So care should be taken when we talk about the
area law of multipartite entanglement and the largest bipartite entanglement.

\section{Entanglement of Gottesman codes and the related codes}

\subsection{Gottesman codes}

A serial quantum codes $[[2^m,2^m-m-2,3]]$ ($m\geq 3$) that fulfill quantum
Hamming bound had been proposed by Gottesman \cite{Gottesman}. By
construction, the first two generators of the stabilizer are $X_1\cdots
X_{2^m}$ and $Z_1\cdots Z_{2^m}.$ An explicit construction of the remaining $%
m$ generators is given by the matrix $(H|CH)$, where $H=\left[
h_0,h_1,...,h_{2^m-1}\right] $ with the $(k+1)^{th}$ column $h_k$ being the
binary vector representing integer $k$ ($k=0,1,...,2^m-1$) and $C$ is any
invertible and fixed point free $m\times m$ matrix, i.e., $Cs\neq 0$ and $%
Cs\neq s$ for all $s$ $\in \Bbb{F}_2^m$ except $s=0.$ The generator $%
Z_1\cdots Z_{2^m}$ is a Z-type generator and omitted hereafter regarding the
entanglement of codewords. We may arrange $H=\left[ H_0,H_1,H_2,\ldots
H_m\right] ,$ with $H_0=h_0=\left[ 0,0,\ldots ,0\right] ^T$, $H_1=\left[
h_{2^{m-1}},h_{2^{m-2}},\ldots ,h_2,h_1\right] =I_{m\times m}$, and $H_j$ is
an $m\times \binom mj$ matrix whose column vector has weight $j.$ The
generator matrix of the last $m$ generators in the standard form will be $%
\left(
\begin{array}{ll}
I & D
\end{array}
\right. \left|
\begin{array}{ll}
F & G
\end{array}
\right) ,$ with $D=\left[
\begin{array}{llll}
H_2 & H_3 & \cdots & H_m
\end{array}
\right] ,F=C,G=CD,$
\begin{equation}
F^T+DG^T=0.  \label{wave8}
\end{equation}
where we have used the facts that any two rows of matrix $H$ are orthogonal,
and each row vector of $H$ has even weight, so $\sum_{i=1}^mH_iH_i^T$ $=0,$
and $\sum_{i=2}^mH_iH_i^T$ $=H_1H_1^T=I.$ Eq. (\ref{wave8}) and the fact
that the generator $X_1...X_{2^m}$ does not contain any $Z$ operator leads
to
\[
\alpha (\mu )=0,
\]
for all $m+1$ dimensional binary vectors $\mu $. To obtain the lower bound
of entanglement for Gottesman codewords, we should verify if condition (\ref
{wave7}) is satisfied or not. When the generator $X_1\cdots X_{2^m}$ is
considered, the whole $D$ matrix for $m+1$ generators is
\[
D=\left[
\begin{array}{llll}
\mathbf{1} & \mathbf{0} & \cdots & (\mathbf{m-1})_{\Bbb{F}_2} \\
H_2 & H_3 & \cdots & H_m
\end{array}
\right] ,
\]
where $\mathbf{1}$ and $\mathbf{0}$ are vectors $(1,1,\ldots ,1)$ and $%
(0,0,\ldots ,0)$ with proper dimensions, respectively. We have
\[
DD^T=\left[
\begin{array}{ll}
\sum_{i=1}^{m^{\prime }}\binom m{2i} & \sum_{i=1}^{m^{\prime }}\mathbf{1}%
H_{2i}^T \\
\sum_{i=1}^{m^{\prime }}H_{2i}\mathbf{1}^T & \sum_{i=2}^mH_iH_i^T
\end{array}
\right] ,
\]
with $m^{\prime }=\left\lfloor m/2\right\rfloor .$ Notice that the $l^{th}$
element of the $m$-dimensional vector $H_{2i}\mathbf{1}^T$ is the weight of $%
l^{th}$ line of $H_{2i}$, each line of $H_{2i}$ has the same weight $t_i$ by
the definition of $H,$ each column of $H_{2i}$ has the same weight $2i,$ so $%
mt_i=2i\binom m{2i}$ is the total weight of the matrix $H_{2i},$ thus $t_i=%
\binom{m-1}{2i-1}$, and $\sum_{i=1}^{m^{\prime }}H_{2i}\mathbf{1}^T=$ $%
\sum_{i=1}^{m^{\prime }}t_i=$ $\sum_{i=1}^{m^{\prime }}\binom{m-1}{2i-1}%
=2^{m-2},$ which is $0$ in $\Bbb{F}_2$ for $m\geq 3,$ so that $%
\sum_{i=1}^{m^{\prime }}H_{2i}\mathbf{1}^T=\mathbf{0}^T$. Meanwhile $%
\sum_{i=1}^{m^{\prime }}\binom m{2i}=\sum_{i=0}^{m^{\prime }}\binom
m{2i}-1=2^{m-1}-1$, which is $1$ in $\Bbb{F}_2$ for $m\geq 2.$ We have
\[
DD^T=I.
\]
The condition (\ref{wave7}) is fulfilled. The entanglement lower bound of
Gottesman codeword is $m+1.$ The number of the generators which are not
Z-type is $m+1,$ so the upper bound of the codeword is $m+1.$ We conclude
that the entanglement of codewords is $m+1$ for Gottesman code $%
[[2^m,2^m-m-2,3]]$ ($m\geq 3$). Written with the length of the code $n=2^m,$
the entanglement of the codewords is
\begin{equation}
E=\log _2n+1.  \label{wave11a}
\end{equation}

\subsection{Family of $8m$ codes \label{sec53}}

The family of codes with parameters $[[8m,8m-l_m-5,3]]$ with $%
l_m=\left\lceil \log _2m\right\rceil $ was constructed\cite{Li}. Alternative
generator matrices of the codes were given \cite{Sixia} based on Gottesman
codes. The number of generators is $l_m+5.$ There is one Z-type generator in
the stabilizer. Thus the upper bound of entanglement (might not be tight) of
codewords is $E_u=l_m+4.$ To obtain the lower bound of entanglement, we will
utilize the generator matrices of \cite{Sixia} directly instead of
transforming them into the standard form of Eq.(\ref{wave3a}). The code can
be divided into $m$ blocks, each block has $8$ qubits. The generator
matrices can be written as
\begin{equation}
\left( A\right. \left| B\right) =\left( A_1,A_2,\ldots ,A_m\right. \left|
B_1,B_2,\ldots ,B_m\right)  \label{wave11b}
\end{equation}
where $A_i$ and $B_i$ are $\left( l_m+5\right) \times 8$ binary matrices,
and every line of $(A_i$ $|B_i)$ is either a line from the generator matrix
of Gottesman [[8,3,3]] code or corresponds to $I^{\otimes 8},X^{\otimes
8},Y^{\otimes 8}$ or $Z^{\otimes 8}$ . It is observed that $A_iA_j^T=0%
\mathbf{,}A_iB_j^T=0\mathbf{,}B_iB_j^T=0\mathbf{,}$ for all $i,j$ . Thus we
have
\begin{equation}
\Gamma =AB^T=\sum_iA_iB_i^T=0\mathbf{.}  \label{wave12}
\end{equation}
Hence $\alpha (\mu )=0\ $for arbitrary binary vector $\mu .$ Meanwhile, we
have $\sum_iA_iA_i^T=0\mathbf{,}$ thus
\begin{equation}
AA^T=0.  \label{wave13}
\end{equation}
Notice that elementary row transformation of $A$ keeps Eq.(\ref{wave13}).
After elementary row transformation, $A$ can be transformed to standard form
\begin{equation}
A\mapsto A^{\prime }=\left[
\begin{array}{ll}
I & D \\
0 & 0
\end{array}
\right] .  \label{wave14}
\end{equation}
Hence
\[
A^{\prime }A^{^{\prime }T}=\left[
\begin{array}{ll}
I+DD^T & 0 \\
0 & 0
\end{array}
\right] =0,
\]
and
\begin{equation}
DD^T=I.  \label{wave15}
\end{equation}
What is crucial is the dimension of the identity matrix in Eq.(\ref{wave14}
). There is one obvious Z-type generator. Besides this one, it is always
possible to work out an full rank identity matrix from $A$ by elementary row
transformation, thus the rank of the identity matrix in Eq.(\ref{wave14}) as
well as in Eq.(\ref{wave15}) is $l_m+4.$ Based on Eq.(\ref{wave12}) and Eq.(%
\ref{wave15}), the lower bound of the entanglement is $E_l=l_m+4$ which
coincides with the upper bound$.$ Hence the entanglement of the codewords of
length $n=8m$ code is
\begin{equation}
E=\left\lceil \log _2n\right\rceil +1.  \label{wave16}
\end{equation}
Notice that Eq. (\ref{wave11a}) for the entanglement of Gottesman codes can
be merged into Eq.(\ref{wave16}).

\subsection{Pasted codes}

The $[[13,7,3]]$ code is obtained \cite{Gottesman96} by pasting Gottesman $%
[[8,3,3]]$ code and cyclic $[[5,1,3]]$ code. For the entanglement of the
codewords of $[[13,7,3]]$ code, there are one Z-type generator among the $6$
generators, the upper bound should be $E_u=5.$ A direct calculation shows
that the lower bound $E_l$ is also $5.$ So we have the entanglement $E=5,$
which fulfill Eq.(\ref{wave16}).

The family of perfect codes $[[n_m,n_m-2m,3]]$ with $n_m=(4^m-1)/3$ for $%
m\geq 3$ is successively constructed by pasting Gottesman $2^{2(m-1)}$ code
(occasionally, we denote the code with its length when it is not confusing)
with $n_{m-1}$ code \cite{Gottesman96} \cite{CRSS}. For the entanglement of
codewords of $n_m$ code, there are $2m$ generators and one of them is Z-type
generator, the upper bound is not difficult found to be $E_u=2m-1$. For the
entanglement lower bound, let's first consider $[[21,15,3]]$ code obtained
by pasting Gottesman $2^4$ code with cyclic $[[5,1,3]]$ code. We may utilize
the special case mentioned at the end of Section \ref{sec4}, then we have $%
E_l=rank_{\Bbb{F}_2}D=5,$ notice that the Z-type generator has already
removed$.$ Similarly, for $n_m$ code, we have $E_l=2m-1.$ The entanglement
is $E=2m-1$ and can be written with respect to the code length $n=n_m$ as
\begin{equation}
E=\left\lceil \log _2n\right\rceil  \label{wave17}
\end{equation}

Another family of codes $[[8n_m,8n_m-2m-3,3]]$ with $m\geq 2$ is
successively constructed by pasting Gottesman $2^{2m+1}$ code with $8n_{m-1}$
code \cite{CRSS}. The entanglement upper bound should be $E_u=2m+2$ which is
determined by the number of non Z-type generators of Gottesman $2^{2m+1}$
code according to the structure of $8n_{m-1}$ code$.$ The entanglement lower
bound is $E_l=2m+2.$ Since the code is a pasting of several Gottesman codes,
the entanglement lower bound can be obtained to be $E_l=2m+2$ by a technic
similar to that of Subsection \ref{sec53}$.$ The entanglement is $E=2m+2$
and can also be written in the form of Eq.(\ref{wave17}) with the code
length $n=8n_m.$

\section{Iteration Algorithm}

Denote $f=\left\langle \overline{0}\right| \left. \Phi _S\right\rangle $,
the closest product state $\left| \Phi _S\right\rangle
=\bigotimes_j(x_j\left| 0\right\rangle +y_j\left| 1\right\rangle )$ with $%
\left| x_j\right| ^2+\left| y_j\right| ^2=1.$ Using Lagrange multiplier
method, we have $L=\left| f\right| ^2-\sum_j\lambda _j(\left| x_j\right|
^2+\left| y_j\right| ^2-1),$ where $\lambda _j$ are the multipliers. The
extremal equations should be $\frac{\partial f}{\partial x_j}f^{*}-\lambda
_jx_j^{*}=0,$ $\frac{\partial f}{\partial y_j}f^{*}-\lambda _jy_j^{*}=0.$
Let $z_j=y_j/x_j,$ we have
\begin{equation}
z_j^{*}=\frac{\partial f/\partial y_j}{\partial f/\partial x_j}.
\label{wave9}
\end{equation}
Thus the group element $M_1^{\mu _1}M_2^{\mu _2}\cdots M_{n-k}^{\mu _{n-k}}$
is isomorphic to $\left( \sum_{i=1}^{n-k}\mu _ia_i\right| \left.
\sum_{i=1}^{n-k}\mu _ib_i\right) =\left( \mu A\right| \left. \mu B\right) ,$
where $\mu =(\mu _1,\mu _2,\ldots ,\mu _{n-k})$ is the binary vector. $\mu A$
and $\mu B$ are binary vectors of length $n.$ So that
\begin{eqnarray}
f &=&N\left\langle 0\right| ^{\otimes
n}\prod_{i=1}^{n-k}(I+M_i)\bigotimes_{j=1}^n(x_j\left| 0\right\rangle
+y_j\left| 1\right\rangle )  \nonumber \\
&=&N\sum_{\mu =\mathbf{0}}^{\mathbf{1}}\left\langle 0\right| ^{\otimes
n}Z^{\mu B}X^{\mu A}(-1)^{\alpha (\mu )}(-i)^{\mu \cdot g}  \nonumber \\
&&\bigotimes_{j=1}^n(x_j\left| 0\right\rangle +y_j\left| 1\right\rangle )
\nonumber \\
&=&N\sum_{\mu =\mathbf{0}}^{\mathbf{1}}(-1)^{\alpha (\mu )}(-i)^{\mu \cdot
g}\left\langle 0\right| ^{\otimes n}X^{\mu A}  \nonumber \\
&&\bigotimes_{j=1}^n(x_j\left| 0\right\rangle +y_j\left| 1\right\rangle )
\nonumber \\
&=&N\sum_{\mu =\mathbf{0}}^{\mathbf{1}}(-1)^{\alpha (\mu )}(-i)^{\mu \cdot
g}\prod_{j=1}^nx_j^{1-(\mu A)_j}y_j^{(\mu A)_j}.
\end{eqnarray}
where $g=(g_1,\ldots ,g_{n-k}),$ and $g_i$ is the number of $Y$ operator in $%
M_i$. From (\ref{wave9}), the iteration equation for $z_j$ is
\begin{equation}
z_j^{*}=\frac{\sum_{\mu |(\mu A)_j=1}(-1)^{\alpha (\mu )}(-i)^{\mu \cdot
g}\prod_{m\neq j}z_m^{(\mu A)_m}}{\sum_{\mu |(\mu A)_j=0}(-1)^{\alpha (\mu
)}(-i)^{\mu \cdot g}\prod_{m\neq j}z_m^{(\mu A)_m}}.
\end{equation}
Notice that the iteration may sometimes fail to reach the global maximum of $%
\left| f\right| ^2$. So, if the ultimate iteration result of the separable
state $\left| \Phi _S\right\rangle $ is the closest product state for $%
\left| \overline{0}\right\rangle ,$ the entanglement of a quantum code from
iteration method will be
\begin{eqnarray*}
E &=&-\log _2\left| f_{*}\right| ^2=n-k-n_s \\
&&-2\log _2\left| \sum_{\mu =\mathbf{0}}^{\mathbf{1}}(-1)^{\alpha (\mu
)}\prod_{j=1}^nx_{j*}^{1-(\mu A)_j}y_{j*}^{(\mu A)_j}\right| .
\end{eqnarray*}
where $n_s$ is the number of Z-type generators, $f_{*},x_{j*}$ and $y_{j*}$
are the extremal values of $f,x_j$ and $y_j$, respectively.

The entanglement for some quantum codes with listed generators by Grassl\cite
{Grassl} is as Table 1. The calculation is based on the iterative algorithm
for unequal upper and lower bounds of entanglement.

\begin{table}[tbp]
\begin{center}
\par
{\bfseries Table 1 The entanglement and the bounds}\\[1ex]
\par
\begin{tabular}{|l||l|l|l|}
\hline
\lbrack[n,k,d]] & E & $E_u$ & $E_l$ \\ \hline
\lbrack [4,1,2]] & 2 & 2 & 2 \\
\lbrack [4,2,2]] & 2 & 2 & 2 \\
\lbrack [5,1,3]] & 2.9275 & 3 & 2 \\
\lbrack [5,2,2]] & 2 & 2 & 2 \\
\lbrack [6,1,3]] & 2.9275 & 3 & 2 \\
\lbrack [6,2,2]] & 3 & 3 & 3 \\
\lbrack [6,3,2]] & 2 & 2 & 2 \\
\lbrack [6,4,2]] & 2 & 2 & 2 \\
\lbrack [7,1,3]] & 3 & 3 & 3 \\
\lbrack [7,2,2]] & 4 & 4 & 3 \\
\lbrack [7,3,2]] & 4 & 4 & 3 \\
\lbrack [7,4,2]] & 3 & 3 & 3 \\
\lbrack [8,1,3]] & 5 & 5 & 4 \\
\lbrack [8,2,3]] & 4.8549 & 5 & 4 \\
\lbrack [8,3,3]] & 5 & 5 & 4 \\
\lbrack [8,4,2]] & 4 & 4 & 4 \\
\lbrack [8,5,2]] & 3 & 3 & 3 \\
\lbrack [8,6,2]] & 2 & 2 & 2 \\
\lbrack [9,1,3]] & 5 & 5 & 4 \\
\lbrack [9,2,3]] & 5 & 5 & 4 \\
\lbrack [9,3,3]] & 5 & 5 & 4 \\
\lbrack [9,4,2]] & 4 & 4 & 4 \\
\lbrack [9,5,2]] & 3 & 3 & 3 \\
\lbrack [9,6,2]] & 2 & 2 & 2 \\ \hline
\end{tabular}
\\[0.5ex]
\end{center}
\end{table}

\section{Conclusions}

The three entanglement measures (the geometric measure, the logarithmic
robustness and the relative entropy of entanglement) are proved to be equal
for quantum stabilizer codeword. The entanglement upper bound of a
stabilizer codeword can be the minimal number of non Z-type generators. A
Z-type generator is the tensor product of identity and/or Pauli $Z$
operators. Further tight upper bound is the ''Pauli persistency'', the
minimal number of Pauli measurements to resolve the entanglement. The
entanglement lower bound based on bipartite entanglement is reduced to a
formula of calculating the ,maximal rank of some matrices. The matrices are
derived from the generator matrix of the code stabilizer. The entanglement
of a self-dual CSS code is proved to be the number of $X$ generators
regardless of the detail structure of the stabilizer. We also derive the
adjacency matrix of the corresponding graph state of CSS code. Upper and
lower bounds of entanglement are given for toric codes, the entanglement is
about half of the code length. Comments are given on the area law of
entanglement for toric codes. The entanglement values of the Gottesman codes
$[[2^m,2^m-m-2,3]]$ ($m\geq 3$), $8m$ codes and Gottesman pasting codes are
equal to their minimal numbers of non Z-type generators. The entanglement $E$
of the Gottesman codes and the related codes scales with the code length $n$
as $E=\left\lceil \log _2n\right\rceil +1$ or $E=\left\lceil \log
_2n\right\rceil +1$. An iterative algorithm is developed to obtain the
entanglement of the codeword as precisely as possible.

\section*{Acknowledgement}

Funding by the National Natural Science Foundation of China (Grant No.
60972071), Zhejiang Province Science and Technology Project (Grant No.
2009C31060)and Natural Science Foundation of Zhejiang Province (Grant No.
Y6100421) are gratefully acknowledged.

\end{document}